\documentclass[preprint,12pt,3p]{elsarticle}
\biboptions{sort&compress}
\usepackage{graphics}
\usepackage{graphicx}
\usepackage{dcolumn}
\usepackage{psfrag}
\usepackage{hyperref}
\usepackage{epsfig,amssymb,amsfonts,amsmath,mathtools,bm,color,xcolor,graphicx,braket,adjustbox,esint,upgreek}

\newcommand{\be}{\begin{equation}}
\newcommand{\ee}{\end{equation}}
\newcommand{\ba}{\begin{eqnarray}}
\newcommand{\ea}{\end{eqnarray}}

\begin{document}

\begin{frontmatter}

\title{\vspace{-1.2cm}
\vspace{5mm}%
Effective range expansion for narrow near-threshold resonances}

\author[1]{Vadim Baru}

\author[2,3]{Xiang-Kun Dong}

\author[4]{Meng-Lin Du}

\author[1]{Arseniy~Filin}

\author[2,3]{Feng-Kun~Guo}

\author[5]{Christoph~Hanhart}

\author[6,7]{Alexey Nefediev}

\author[4]{Juan Nieves }

\author[8,9,10]{Qian Wang}

{\address[1]{Institut f\"ur Theoretische Physik II, Ruhr-Universit\"at Bochum, D-44780 Bochum, Germany }}

{\address[2]{CAS Key Laboratory of Theoretical Physics, Institute of Theoretical Physics, Chinese Academy of Sciences, Zhong Guan Cun East Street 55, Beijing 100190, China}}

{\address[3]{School of Physical Sciences, University of Chinese Academy of Sciences, Beijing 100049, China}}

{\address[4]{Instituto de F\'{\i}sica Corpuscular (centro mixto CSIC-UV), Institutos de Investigaci\'on de Paterna, C/Catedr\'atico Jos\'e Beltr\'an 2, E-46980 Paterna, Valencia, Spain}}

{\address[5]{Institute for Advanced Simulation, Institut f\"ur Kernphysik and J\"ulich Center for Hadron Physics, Forschungszentrum J\"ulich, D-52425 J\"ulich, Germany}}

{\address[6]{P.N. Lebedev Physical Institute of the Russian Academy of Sciences, 119991, Leninskiy Prospect 53, Moscow, Russia}}

{\address[7]{Moscow Institute of Physics and Technology, 141700, Institutsky lane 9, Dolgoprudny, Moscow Region, Russia}}

{\address[8]{{Guangdong Provincial Key Laboratory of Nuclear Science,}\\ Institute of Quantum Matter, South China Normal University, Guangzhou 510006, China}}

{\address[9]{Institute of High Energy Physics, Chinese Academy of Sciences, Beijing 100049, China}}

{\address[10]{Guangdong-Hong Kong Joint Laboratory of Quantum Matter,\\Southern Nuclear Science Computing Center, South China Normal University, Guangzhou 510006, China}}

\begin{abstract}
We discuss some general features of the effective range expansion, the content of
its parameters with respect to the nature of the pertinent near-threshold states and
the necessary modifications in the presence of coupled channels, isospin violations and unstable 
constituents.  As illustrative examples, we analyse the properties of the $\chi_{c1}(3872)$ and $T_{cc}^+$ states
supporting the claim that these exotic states have a predominantly molecular nature.
\end{abstract}

\end{frontmatter}

\section{Introduction}
\label{Intro}

The experimental progress in the last two decades led to the discovery of many states in the charmonium and bottomonium mass range that are in conflict with the naive quark-model picture, for recent reviews see, for example, Refs.~\cite{Chen:2016qju,Lebed:2016hpi,Esposito:2016noz,Guo:2017jvc,Olsen:2017bmm,Ali:2019roi,Brambilla:2019esw,Guo:2019twa}. Among these exotic states, the first unconventional quarkonium-like one, the $\chi_{c1}(3872)$, also known as $X(3872)$, discovered more than a decade ago by the Belle Collaboration~\cite{Belle:2003nnu} remains one of the most popular and controversial states in theoretical and experimental studies. Since this shallow state is located less than 100 keV below the $D^0\bar D^{*0}$ threshold, 
it appears as a natural candidate for a hadronic molecule, which was predicted in Ref.~\cite{Tornqvist:1993ng}, though other interpretations, such as a tetraquark 
or a mixture of a charmonium state with a meson molecule 
are also possible (see the review articles above for the original publications). Here it should be clear that what is discussed is the leading contribution; in particular, the molecular component of a given state shows up as the importance of the long-distance parts of the wave function and at short distances other contributions could mix in. Very recently, the LHCb Collaboration announced the discovery of an isoscalar double-charm exotic state $T_{cc}^+$, whose quantum numbers are favoured to be $J^P=1^+$. It is located just about 360 keV
below the $D^0D^{*+}$ threshold~\cite{LHCb:2021vvq,LHCb:2021auc}, which makes it a very close analogue of the $\chi_{c1}(3872)$ in the spectrum of states containing a $c\bar{c}$ pair. 

To shed light on the nature of near-threshold states and discriminate between molecular and compact near-threshold ones, Morgan proposed the concept of pole counting as a tool for the classification of near-threshold shallow states in the one-channel problem~\cite{Morgan:1992ge}. 
Specifically, the existence of only one pole in the complex (momentum) $k$-plane near $k$=0 points towards a molecular scenario. 
On the contrary, the appearance of a pair of poles in the complex $k$-plane in the vicinity of the threshold corresponds to the case where a resonance has a large admixture of a compact state. The pole counting in the weak binding limit contains the same information as the renormalisation factor $Z$ used by Weinberg in Ref.~\cite{Weinberg:1965zz} to study the properties of the deuteron as was demonstrated in Ref.~\cite{Baru:2003qq}. 
Both approaches require the knowledge of the pertinent information about the scattering problem contained at low energies in the effective range parameters of the scattering amplitude.


It is a direct consequence of unitarity that the imaginary part of the inverse single-channel two-particle scattering amplitude $f$
in the non-relativistic kinematics is given by
\begin{equation}
\Im\left(f(E)^{-1}\right) = -\frac{2\pi}{\mu} \, \Im\left(T(E)^{-1}\right) = -k \ ,
\label{eq:imf}
\end{equation}
where $T$ denotes the $T$ matrix, $k=\sqrt{2\mu E}$, $\mu$ stands for the reduced mass of the scattering particles 
and $E$ is the energy of the system relative to the threshold. On the other hand, 
it was demonstrated long ago that the real part of the inverse scattering amplitude is a polynomial in the variable $E$ or
equivalently in even powers of $k$~\cite{Blatt:1949zz, Bethe:1949yr} and, in the case of $S$ wave, it can be written as
\begin{equation}
\Re\left(f(E)^{-1}\right) =-\frac{2\pi}{\mu} \, \Re\left(T(E)^{-1}\right) = k \cot\delta= \frac{1}{a}+\frac12 r k^2+{\cal O}(k^4) \ ,
\label{eq:ref}
\end{equation}
where the parameters $a$ and $r$ are called the scattering length and effective range, respectively. 
The given expression holds for regular potentials of a finite range and calls for modifications in the presence of singular potentials.\footnote{ In the presence of a long-range singular potential such as the Coulomb-type interaction, the ERE, as given in Eq.~\eqref{eq:ref}, can not be applied and requires modifications which go beyond the scope of this work. }
Note
that different sign conventions can be used for the scattering length.
The one employed here is common in particle physics and refers to the case, when a mildly attractive interaction
leads to a positive scattering length. Meanwhile, a repulsive interaction, which does not yield any composite states, or a strongly attractive one yielding a bound state 
corresponds to a negative scattering length. 
An opposite sign convention of the $1/a$ term in Eq.~(\ref{eq:ref}) is commonly adopted in nuclear physics and also employed in Refs.~\cite{Blatt:1949zz, Bethe:1949yr}.

Weinberg related the parameters of the effective range expansion (ERE) to $Z$, the probability to find the 
compact component of a given hadron inside a bound state wave function~\cite{Weinberg:1965zz},
\begin{equation}\label{eq:Wein}
a = -2\left(\frac{1-Z}{2-Z}\right)\frac{1}{\gamma} + {\cal O}\left(\frac{1}{\beta}\right) \quad
\mbox{and} \quad r = -\left(\frac{Z}{1-Z}\right)\frac{1}{\gamma} + {\cal O}\left(\frac{1}{\beta}\right) ,
\end{equation}
where $\gamma=\sqrt{2\mu |E_b|}$ (with $E_b<0$ for the binding energy) is the binding momentum and $1/\beta$ estimates range corrections. Since $\beta$ denotes the next momentum scale that is not treated explicitly in the ERE,  it is normally estimated as the mass of the lightest exchange particle~\cite{Voloshin:1976ap,Tornqvist:1993ng}.
However, it may also be the momentum scale due to the presence of the next closed channel --- to be discussed below. Clearly, model-independent statements are possible only if $\gamma \ll \beta$.
Then one observes that 
\begin{eqnarray}\label{eq:mol}
a\to -\frac{1}{\gamma} \ &\&& \ r\to \phantom{-}\frac{N_r}{\beta} \quad \mbox{for} \ Z\to 0 \quad \mbox{(predominantly molecular)} \ , \\
\label{eq:comp}
a\to -\frac{N_a}{\beta} \ &\&& \ r\to -\infty \quad \mbox{for} \ Z\to 1 \quad \mbox{(predominantly compact)} \ ,
\end{eqnarray}
where $N_a$ is expected to be a positive number of the order of 1. In case of a single-channel potential scattering with a finite interaction range and negative potential in the whole space,\footnote{The condition of a negative-definite potential in the whole coordinate space does not hold when there is a repulsive barrier in addition to the attractive part of the potential. In that case, there can be a bound state and at the same time a negative effective range.} also $N_r$ should be
positive and of the order of 1, as demonstrated long ago by Smorodinsky~\cite{Smorodinsky, Landau:1991wop}, see also Ref.~\cite{Esposito:2021vhu} for a recent discussion.
However, this conclusion does not hold when coupled channels are included, as we discuss below.

Interest in Weinberg's analysis, for a long time applied to the deuteron only, was revived,
when it was demonstrated in Ref.~\cite{Baru:2003qq} that the same kind of analysis can also be used for unstable states as long as the inelastic threshold is sufficiently remote. Later various attempts were made to generalise the scheme to coupled channels and resonances~\cite{Gamermann:2009uq,Sekihara:2016xnq,Sekihara:2015gvw,Kamiya:2015aea,Kamiya:2016oao,Aceti:2012dd,Aceti:2014ala,Guo:2016wpy,Guo:2015daa,Hyodo:2013nka,Hyodo:2011qc,Kang:2016ezb,Oller:2017alp,Sekihara:2014kya,Xiao:2016wbs,Xiao:2016dsx,Hyodo:2013iga,Guo:2020pvt} and also to virtual states~\cite{Matuschek:2020gqe}.
For a detailed recent discussion of issues in applying the Weinberg criterion to the case of a positive effective range, we refer to Ref.~\cite{Li:2021cue}.
Last but not least, the analysis discussed above implicitly relies on the assumption that the scattering amplitude does not have Castillejo-Dalitz-Dyson (CDD) zeros in the vicinity of the threshold, since
such zeros would invalidate the ERE given in Eq.~\eqref{eq:ref} and thus require a more advanced analysis of the near-threshold states --- for the corresponding extensions of the original analysis the interested reader is referred to Refs.~\cite{Baru:2010ww,Kang:2016jxw}. 
Note, however, that the LHCb data for the $\chi_{c1}(3872)$ \cite{LHCb:2020xds} and $T_{cc}^+$ \cite{LHCb:2021vvq,LHCb:2021auc} states
discussed in the manuscript do not provide evidence for the near-threshold CDD zeros and are completely consistent with the description within the ERE. Thus, until an experimental evidence of such additional zeros is obtained, we employ the Occam’s razor principle to assume that the ERE as given in Eq.~\eqref{eq:ref} is appropriate for the systems to be considered.

In this work we investigate a series of issues related to the ERE as well as its application
to extract information on the structure of near-threshold states. Hence, we discuss/provide
\begin{itemize}
\item[$\bullet$] a generalisation to coupled channels, that allows to make contact to the usual parameterisations of near-threshold
resonance states (Sec.~\ref{sec:cc});
\item[$\bullet$] the role of finite range corrections (Sec.~\ref{sec:finiterange});
\item[$\bullet$] the role of isospin violation --- this is of particular interest for the exotic states $\chi_{c1}(3872)$ and $T_{cc}^+$ with their poles very close to the $D^0\bar D^{* 0}$ and
 $D^0D^{* +}$ thresholds, respectively (Sec.~\ref{sec:IV});
\item[$\bullet$] insights on the nature of the $\chi_{c1}(3872)$ and $T_{cc}^+$ states from the Weinberg compositeness criterion (Sec.~\ref{sec:compos});
\item[$\bullet$] a way to account for the finite width of the pertinent scattering states (Sec.~\ref{sec:finitewidth}).
\end{itemize}
Some, but not all, of those items have already been addressed in the literature, as will also become clear in the
discussion below. However, given the renewed interest
in exploiting the ERE to access information about exotic states and to gain insights into their nature,
 we regard it both timely and important to provide a concise and comprehensive overview
over the properties of the ERE. 
What will be left for future research is the possible role of the pion exchange and, in particular, of nearby three-body
thresholds as well as a possible coupling to $D$ waves.  Specifically, the inclusion of tensor interactions from the one-pion exchange may generate a potential barrier and, in this way, shift the poles from bound/virtual states to resonances. 
Nevertheless, we do not expect such a barrier to affect the nature of the states. 
An example of such a situation is provided by the $Z_b(10610)$ and $Z_b(10650)$ which become resonance states, as soon as tensor forces are included, but still remain molecules 
\cite{Wang:2018jlv,Baru:2019xnh}.

\section{Role of coupled channels}
\label{sec:cc}

In Ref.~\cite{Hanhart:2007yq}, a scheme was developed that allows for a combined analysis of the various decay
channels of the $\chi_{c1}(3872)$. The starting point is the coupled-channel $S$-wave scattering amplitude of two hadrons, 
which in the presence of a nearby pole
 and the absence of additional zeros discussed in the Introduction can be written as~\cite{Baru:2003qq, Baru:2010ww, Hanhart:2011jz}
\begin{equation}\label{Eq:Flatte}
{ f_{ab}(E) = - \frac{g_a g_b}{2D(E)}} \, ,
\end{equation}
with $D(E)$ given by
\begin{equation}
D(E) = E-E_f+\frac{i}{2}\left(g_1^2 k_1 + g_2^2 k_2 + \sum_i\Gamma_i(E)\right) ,
\label{eq:DEorig}
\end{equation}
where the energy dependence of the two elastic (neutral $D^0 \bar D^{*0}$ and charged $D^\pm D^{*\mp}$) channels\footnote{The appropriate linear combination of the relevant channels corresponding to the positive $C$ parity state is implied.} is explicitly kept. In addition, $g_1$ and $g_2$ are the effective couplings of the neutral and charged channels to the $\chi_{c1}(3872)$, respectively. For real values of the energy $E$, analyticity  in the physical Riemann sheet requires that
\begin{equation}
k_a = \sqrt{2\mu_a (E-\delta_a)}\Theta(E-\delta_a) + i\sqrt{2\mu_a (\delta_a-E)}\Theta(\delta_a-E) \, .
\end{equation}
In what follows, we will define the energy $E$ relative to the lowest threshold, which we assign to channel
1. Accordingly, we set 
$$
\delta_1 =0 \qquad \mbox{and} \qquad \delta_2 = m_1^{(2)} + m_2^{(2)} - m_1^{(1)} - m_2^{(1)} \, , 
$$
where $m_k^{(a)}$ denotes the mass of the $k^{\rm th}$ particle in the elastic channel $a$  and $\mu_a$ is the corresponding reduced mass.
The terms $\Gamma_i(E)$ are meant to absorb all other inelasticities. Explicit forms for those in the case of the 
$\chi_{c1}(3872)$ decays into $J/\psi \omega$ and $J/\psi \rho$ are given in Ref.~\cite{Hanhart:2007yq}.
Note that the presence of such terms does not lead to a violation of unitarity~\cite{Hanhart:2015cua}.
Then, for the production rates in various possible elastic ($a$) and inelastic ($i$) final states one has~\cite{Hanhart:2015cua}
\begin{eqnarray}
\frac{d\mbox{Br}(S\to a+...)}{dE} = {\cal N}\frac{g_a^2 k_a}{|D(E)|^2} \qquad \mbox{and} \qquad 
\frac{d\mbox{Br}(S\to i+...)}{dE} = {\cal N}\frac{\Gamma_i(E)}{|D(E)|^2} \ ,
\label{eq:spectra}
\end{eqnarray}
where $S$ denotes some source, the ellipsis allows for the presence of additional particles in the final states, and
all factors common to the different transitions are absorbed in the source-specific prefactor ${\cal N}$.
The given expressions are valid under a reasonable assumption that the inelastic channels are not essential for the production of the $\chi_{c1}(3872)$~\cite{Dong:2020hxe}.
To simplify the discussion, in what follows, we assume that the energy dependence of the terms $\Gamma_i(E)$ from inelastic channels
is negligible and thus the sum over all of them can be absorbed into a single constant $\Gamma_{\rm inel.}$.

When Eq.~(\ref{eq:spectra}) is used in fits to data with a pole located at slightly negative values
of $E$, there appears a problem, as pointed out in Ref.~\cite{LHCb:2020xds}. Namely, there exists a very strong correlation between
the coupling $g_1^2$ and the mass parameter $E_f$. That such correlations appear is quite natural, since the parameterisation in Eq.~\eqref{Eq:Flatte} leads to $E_p$, the real part of the pole location in the physical Riemann sheet (bound state), given by
\begin{equation}
E_p = E_f +\frac{1}{2} (g_1^2 \gamma_1 + g_2^2 \gamma_2) \, ,
\label{eq:correlation}
\end{equation}
where $\gamma_a= \sqrt{2\mu_a(\delta_a-E_p)}$ ($a=1,2$) are both real-valued and positive as long as $E_p<0$. Clearly, only
$E_p$ is observable and, accordingly, changes in $E_f$ can be traded for changes of $g_1$ and/or $g_2$
(in the isospin limit assuming the $\chi_{c1}(3872)$ to be an isoscalar, one finds $g_1^2 =g_2^2$, which was used 
in the analyses of Refs.~\cite{Hanhart:2007yq,LHCb:2020xds}).
To remove this correlation we, therefore, propose to replace Eq.~(\ref{eq:DEorig}) by
\begin{equation}
D(E) = E-E_p+\frac{i}{2}\left(g_1^2 (k_1-i\gamma_1) + g_2^2 (k_2-i\gamma_2) + \sum_i\Gamma_i(E)\right) .
\label{eq:DE}
\end{equation}
This expression is equivalent to the original one, except that now $E_p$ is directly the real
part of the pole location while its imaginary part is provided by the last term in parentheses. 
If the pole is located on the unphysical Riemann sheet of the complex energy plane with respect to the lower threshold, the
term $(k_1-i\gamma_1)$ needs to be replaced by $(k_1+i\gamma_1)$. 
Since we search for the pole near the lowest threshold, the momentum of the upper channel is on its physical sheet.
From here on, for the sake of definiteness, we assume the $\chi_{c1}(3872)$  to emerge from
a pole on the physical sheet (conventionally called bound state, without
implying anything on the nature of the state) and proceed with the discussion based on Eq.~(\ref{eq:DE}).
Then $\gamma_1$ refers to the binding momentum with respect to the lower threshold.  It should be noted, however, that the analysis performed is insensitive to the sign of $\gamma_1$, so that its generalisation to a virtual state is quite straightforward.

The first diagonal element of the Flatt\' e parameterisation (Eqs.~\eqref{Eq:Flatte} and \eqref{eq:DE}), which corresponds to the lowest-threshold channel, can now be straightforwardly used to determine the 
ERE parameters of the coupled-channel scattering amplitude. Then, for the energies in the
proximity of the lower threshold, using $E=k_1^2/(2\mu_1)$ and
\begin{equation}
k_2 = i\sqrt{2\mu_2\left(\delta_2-\frac{k_1^2}{2\mu_1}\right)} = i\sqrt{2\mu_2\delta_2} - \frac{i}2\sqrt{\frac{\mu_2}{2\mu_1^2\delta_2}} \ k_1^2 + {\cal O}\left(\frac{k_1^4}{\mu_1^2\delta_2^2}\right),
\label{eq:k2expand}
\end{equation}
one finds for the scattering length and effective range, respectively, 
\begin{equation}
a = -\frac{g_1^2}{\gamma_1^2/\mu_1+g_1^2\gamma_1+g_2^2(\gamma_2-\sqrt{2\mu_2\delta_2})+i\Gamma_{\rm inel.}}, \quad r=-\frac{2}{\mu_1 g_1^2} - \frac{g_2^2}{g_1^2}\sqrt{\frac{\mu_2}{2\mu_1^2\delta_2}} \ .
\label{eq:arcc}
\end{equation} 
The expressions \eqref{eq:arcc} agree to those originally provided in Ref.~\cite{Matuschek:2020gqe} for the $ DK/D_s\eta$ coupled-channel problem in the context of the $D_{s0}^*(2317)$ state and reduce to those of Ref.~\cite{Esposito:2021vhu}
when the above mentioned isospin relation,  $g_1^2=g_2^2$, is employed. 

Weinberg's criterion associates a negative effective range, whose modulus is large compared to the range of the forces, to a predominantly
compact state.

The authors of Ref.~\cite{Esposito:2021vhu} take $g_1^2=0.108\pm 0.003$ 
(with $E_f$ fixed to $-7.2$ MeV) from the LHCb analysis of the $\chi_{c1}(3872)$ line shape~\cite{LHCb:2020xds} to obtain 
\begin{equation}
-r_{\rm LHCb} = 5.34 \ \mbox{fm} \gg 1/M_\pi=1.43\ \mbox{fm}, \label{eq:rLHCb}
\end{equation}
with $M_\pi$ for the pion mass, and use this result to conclude that the exotic $\chi_{c1}(3872)$ must be predominantly a compact state. 
However, two comments are here in order.
First, it is necessary to take into account that the second term in the expression for the effective range $r$
 in Eq.~(\ref{eq:arcc}) stems from coupled-channel dynamics and clearly needs to be attributed to the molecular component of the $\chi_{c1}(3872)$ --- we come back
 to this point below in Sec.~\ref{sec:IV}. Accordingly, what should be compared with the range of forces
 in the spirit of the Weinberg's criterion is not $r_{\rm LHCb}$, but
 \begin{equation}
 r_{\rm LHCb} + \sqrt{\frac{\mu_2}{2\mu_1^2\delta_2}} =-\frac{2}{\mu_1 g_1^2} = -3.78 \ \mbox{fm} \, .
 \label{eq:rcorr}
 \end{equation}
 
\begin{figure}[t]
\centerline{\includegraphics[width=0.7\textwidth]{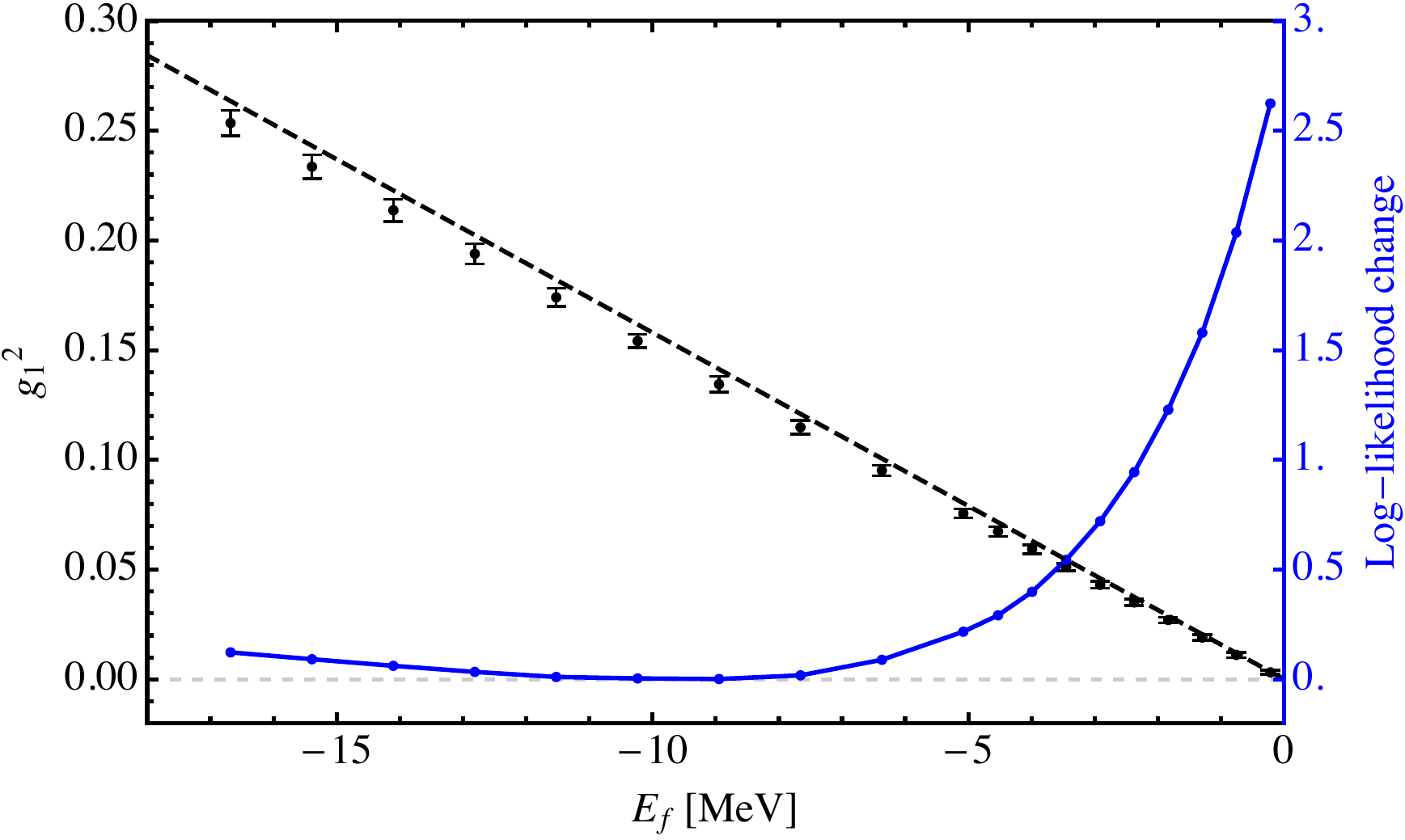}}
\caption{Coupling (black dots with errors) to the elastic $D^0\bar D^{*0}$ channel ($g_1^2$) and change in the negative log-likelihood function relative to its minimum (blue dots) as function of the Flatt\'e energy $E_f$ (Eq.~(\ref{eq:DEorig})). Results are taken from the LHCb study~\cite{LHCb:2020xds} of the line shape of the $\chi_{c1}(3872)$ state.
The black dashed line shows the approximate result from Eq.~\eqref{Eq:g1sq}.
 \label{fig:LHCb} }
\end{figure} 
 
Furthermore, due to the large correlation between the coupling and mass parameter discussed above in Eq.~(\ref{eq:correlation}), in the LHCb analysis of Ref.~\cite{LHCb:2020xds}, the value of the mass parameter $E_f$ was fixed to $-7.2$ MeV. Thus, 
the uncertainty of the coupling related to the variation of $E_f$ was not taken into account in the value of $g_1^2=0.108\pm 0.003$ used in Ref.~\cite{Esposito:2021vhu} to obtain $r_{\rm LHCb}$ quoted in Eq.~\eqref{eq:rLHCb}. 
Results from the LHCb study~\cite{LHCb:2020xds} of the line shape of the $\chi_{c1}(3872)$ state are shown in Fig.~\ref{fig:LHCb} and allow two instructive conclusions. On the one hand, we note that the minimum of the log-likelihood function is not exactly located at $E_f=-7.2$ MeV, but rather it is shifted to the left and found at $E_f\simeq-9$ MeV, for which $g_1^2$ is already around 25\% larger. Though this variation of the central value of $E_f$ hardly changes the binding energy, which remains around 21 keV, it affects noticeably the effective range, so that instead of Eq.~\eqref{eq:rcorr} one would rather get
\be
r_{\rm LHCb} + \sqrt{\frac{\mu_2}{2\mu_1^2\delta_2}} =-\frac{2}{\mu_1 g_1^2} \approx -3 \ \mbox{fm}.
\label{rLHCbnew}
\ee

On the other hand, as discussed in Ref.~\cite{LHCb:2020xds}, and also illustrated here in Fig.~\ref{fig:LHCb}, the minimum is very shallow and the negative log-likelihood relative to its minimum value ($\Delta LL$) rises very slowly, with lower values of $E_f$ being counterbalanced with a linear increase in the coupling to the $D\bar D^*$ channels. Actually, $\Delta LL$ is not increased by one unit up to huge values of $ E_f $ around -270 MeV. However, values of $E_f$ approaching the $D\bar D^{*0}$ threshold are disfavoured, with good quality fits obtained only for negative values of $E_f$.

As a consequence, much larger 
values of $g_1^2$ appear to be consistent with the change of the log-likelihood function by unity. 
In fact, the black symbols in Fig.~\ref{fig:LHCb} follow nicely the pattern predicted
from Eq.~(\ref{eq:correlation}), 
\begin{equation}\label{Eq:g1sq}
g_1^2 \approx -\sqrt{\frac{2}{\mu_2\delta_2}} \ E_f= -\frac{0.0158}{\mbox{MeV}} \ E_f \ ,
\end{equation}
(see the dashed line in Fig.~\ref{fig:LHCb}) where we used  that $g_1^2=g_2^2$ in the isospin limit and that for $|E_f|$ of the order of 10 MeV or larger the effects of a non-vanishing pole mass $|E_p|<0.1$ MeV 
can be safely neglected. Therefore, based on these considerations the coupling $g_1^2$ deduced in Ref.~\cite{LHCb:2020xds}
should be regarded as a lower bound and accordingly the absolute value of the effective range 
given in Eq.~(\ref{eq:rcorr}) as an
upper bound. For instance, taking $E_f = -270$~MeV one would naturally get for the effective range as small value as $-0.1$~fm, which is already consistent with 0 given the accuracy of the approach. 
That the analysis by LHCb with its current mass resolution is not sensitive to the range corrections is nicely illustrated in Fig.~4 of Ref.~\cite{LHCb:2020xds}. Indeed, 
as long as the energy resolution function is not included, the shape of the Flatt\' e distribution is clearly asymmetric
(see the red curve in the left panel of that figure) which means that this distribution allows one at least in principle\footnote{
It is shown in Ref.~\cite{Baru:2004xg} that for a large coupling to the hadronic channel, the Flatt\' e distribution in the near-threshold region shows a scaling behaviour which 
does not allow one to determine all the parameters individually, but rather their ratios.} to extract the coupling to the elastic channel $g_1^2$. However, once the Flatt\' e distribution is convolved with the energy resolution, the resulting shape appears to be simply indistinguishable from the Breit-Wigner distribution. This is shown in the right panel in Fig.~4 of Ref.~\cite{LHCb:2020xds}.

Based on this logic we can safely conclude that the $\chi_{c1}(3872)$ line shape is completely consistent
with a purely molecular nature of it.

Recently, LHCb reported a similar analysis of the $T_{cc}^+$ state \cite{LHCb:2021auc}, located right below the $D^0D^{*+}$ threshold,  from which the scattering length was extracted to be
\be \label{Eq:aTcc}
a = [ -(7.16 \pm 0.51) + i (1.85 \pm 0.28)] \ \mbox{fm} .
\ee %
In this case, however, 
only an upper bound on the absolute value of the negative effective range was found to be 
\be \label{Eq:rTcc}
0 \le |r| < 11.9\, (16.9) \, \mbox{fm at 90\% (95\%) CL}, 
\ee
with its value consistent with 0 for the baseline fit. With a large scattering length, as given above, 
this state is also consistent with a hadronic molecule, as discussed below. 

\section{Finite range corrections}
\label{sec:finiterange}

As follows from Eq.~\eqref{eq:arcc}, in the Weinberg analysis effective ranges are always negative. One way to understand
this is to interpret the expressions in terms of an effective field theory with point-like interactions where the binding momentum $\gamma_1$ is retained as the
only dynamical scale, and all finite range corrections are integrated out. On the other hand, Wigner has
shown long time ago that effective ranges should not exceed the range of forces for otherwise
causality would be violated~\cite{Wigner:1955zz}; for a simple derivation of the Wigner bound we refer to Appendix A of Ref.~\cite{Matuschek:2020gqe}, see also Ref.~\cite{Hammer:2010fw}
for a related discussion.
Therefore, in an effective theory with zero range interactions the effective ranges need to be negative.
The easiest way to go beyond the point-like approximation is to retain in the construction of the
function $D(E)$ the dispersive corrections to the $k_1$ and $k_2$ terms. To that end we consider the hadronic loop,
\be
J_a(k_a) = \frac{2}{\pi}\int \frac{f_a^2(q) q^2}{q^2- k_a^2 -i 0} dq \,,
\ee
where $f_a(q)$ is the vertex function normalised such that $f_a(k_a)=1$. The real parts of the loop functions provide range corrections in Eq.~\eqref{eq:DEorig}, which are taken into account by replacing in that equation
\be
\frac{i}{2} (g_1^2 k_1 + g_2^2 k_2) \longrightarrow \frac12 g_1^2 \left[J_1(k_1)-J_1(0)\right]+\frac12 g_2^2 \left[J_2(k_2)-J_2(0)\right].
\ee
The finite-range contribution to the effective range reads
\begin{align*}
r_{\rm finite}& \equiv -2 \frac{\partial (J_1-ik_1)}{\partial k_1^2}\bigg|_{k_1^2=0} -2\frac{g_2^2}{g_1^2} \frac{\partial (J_2-ik_2)}{\partial k_2^2}\frac{\partial k_2^2}{\partial k_1^2}\bigg|_{k_1^2=0}
\\
& = -2 \frac{\partial (J_1-ik_1)}{\partial k_1^2}\bigg|_{k_1^2=0} -2\frac{\mu_2g_2^2}{\mu_1g_1^2} \frac{\partial (J_2-ik_2)}{\partial k_2^2}\bigg|_{k_1^2=0}.
\end{align*}
In particular, in a point-like theory with pions integrated out, choosing $f_a^2(q) = \theta(\Lambda-q)$ with $\Lambda\sim M_{\pi}$ 
yields
\be
r_{\rm finite}= \left(1 +\frac{{\mu_2}g_2^2}{{\mu_1}g_1^2}\right) \frac{4}{\pi\Lambda} \,.
\ee
Then, if there is only one channel ($g_2=0$), $r_{\rm finite}\simeq {4}/({\pi M_{\pi}}) = 1.8$ fm. This is the typical size of the effective range in the deuteron case, which is controlled by the left-hand cut contribution from the one-pion exchange (OPE) potential, see also Ref.~\cite{Baru:2015ira} for a related discussion. The OPE in the $\chi_{c1}(3872)$ and $T_{cc}^+$ systems has a subtlety because of a three-body cut which opens a few MeV below the corresponding resonance pole \cite{Baru:2011rs,Du:2021zzh}. On the one hand, it formally sets the range scale $\Lambda\sim\sqrt{2M_\pi(m_{D^*}-m_D-M_\pi)}\simeq 40$~MeV and thus one might naively expect the values for $r_{\rm finite}$ around 3 times larger than given above for $\Lambda\simeq M_\pi$. On the other hand, there are good reasons to expect that the OPE contribution to the effective range here is suppressed relative to the $NN$ case.
First, the coupling constant of the pion with the heavy mesons is strongly suppressed relative to that with the $NN$ system, as discussed in Ref.~\cite{Fleming:2007rp}. 
Also, 
the small momenta generated from the cut
provide an additional suppression. 
Therefore, the leading contribution to the range is expected to come from shorter-range contributions. 
Then, taking $\Lambda\sim 500$ MeV and $g_1^2=g_2^2$, the finite-range contribution to the effective range of the $\chi_{c1}(3872)$ and $T_{cc}^+$ can be estimated as
\be
r_{\rm finite}\simeq 1~\mbox{fm}. 
\label{rfin}
\ee

\section{Isospin violation}
\label{sec:IV}

As discussed in Sec.~\ref{Intro} for the $\chi_{c1}(3872)$, the correction to the effective range from the isospin violating term is $-1.57$~fm (see Eq.~\eqref{eq:rcorr}), while for the $T_{cc}^+$ it is even larger, namely, $-3.78$~fm. In this section we discuss these corrections in more detail.

Both the $\chi_{c1}(3872)$ and $T_{cc}^+$ are assumed to be isoscalar states. To implement this in 
the expressions for the line shapes, one can choose  $g_1^2=g_2^2$ and only retain the leading source of isospin
violation explicitly, which is provided by the splitting in the thresholds of the different charge states,
called $\delta_2$ above (in case of the $\chi_{c1}(3872)$ there is also isospin violation that becomes
visible in the coupling to the inelastic channels but this is not of interest here).
Accordingly, isospin symmetry is recovered for $\delta_2 \to 0$. However, in this limit one gets from
Eq.~(\ref{eq:arcc}) that $r\to -\infty$, which clearly cannot make sense. The problem here is that the expansion
of $k_2$ provided in Eq.~(\ref{eq:k2expand}) is justified only if $k_1^2\sim \gamma_1^2\ll 2\mu_1\delta_2$.
In the mentioned isospin limit, however, this inequality clearly does not hold. 
Therefore, as long as we deal with a state located near the elastic threshold 1 and obeying $|E_P| = \gamma_1^2/(2 \mu_1) \ll \delta_2$, we propose the following method 
to relate the ERE parameters (the scattering length and effective range) extracted from a fit to experimental data employing Eq.~(\ref{eq:DE}) 
to the ones to be used in the Weinberg criterion: 
\begin{enumerate}
\item[1)] subtract from $a^{-1}$ and $r$ extracted from Eq.~(\ref{eq:DE}) all isospin-symmetry-violating terms related with the upper channel 2 and supply $r$ with all finite-range corrections;
\item[2)] if there is an inelastic channel with $\delta_{\rm inel}\gg |E_P|$ (here $\delta_{\rm inel}$ is the energy distance between the thresholds of the elastic channel 1 and inelastic one), subtract from $a^{-1}$ and $r$ all coupled-channel effects related to this inelastic channel.
\end{enumerate}

This ways one removes all hadronic coupled-channel effects and arrives at the scattering length and effective range from channel 1 to be used in the Weinberg criterion,
\begin{equation}\label{Eq:ERE1ch}
a=-\frac{1}{\gamma_1}\left(1+\frac{\gamma_1}{\mu_1 g_1^2}\right)^{-1}, 
\quad r=-\frac{2}{\mu_1 g_1^2} \ + \ r_{\rm finite} \, .
\end{equation}

 We note also that the correction from channel 2 to the effective range is by far more important than that to the scattering length. The dominant contribution to the scattering length comes from the binding momentum $\gamma_1$, while the correction from the second channel is parametrically suppressed as $\sqrt{|E_P|/\delta_2}$ relative to the leading term. Therefore, as long as $|E_P|$ is much smaller than $\delta_2$, the correction from the upper channel to the scattering length is suppressed.

 As an alternative to the method proposed above, one could in principle first switch to the isospin limit and then apply the Weinberg criterion. 
However, while this can be implemented straightforwardly for the effective range, quantifying the impact of this limit on the pole and therefore on the scattering length is possible only when a dynamical equation is solved for the state of interest as it is described, e.g., for the $T_{cc}^+$ in Refs.~\cite{Albaladejo:2021vln,Du:2021zzh}. 
Indeed, the pole position of the state in the isospin limit with the isospin-averaged meson masses, which is a possible convention, may differ from the one extracted from experiment. 
 This would force also the scattering length to change in a nontrivial way which could only be inferred from solving a dynamical problem. 
 Thus, if one needs to extract the nature of the state from the ERE directly, without solving a dynamical problem, the method proposed above is preferred. 
 
\section{Weinberg criterion of compositeness}
\label{sec:compos}

In Ref.~\cite{Matuschek:2020gqe}, a generalisation of the Weinberg criterion is introduced to characterise the compositeness ($\bar X$) of bound, virtual and resonance states. It reads
\be\label{ex:barX}
\bar X= \sqrt{\frac{1}{1+|2r/ a|} }.
\ee
The original Weinberg criterion 
could not be applied to systems having bound states with positive effective ranges~\cite{Matuschek:2020gqe} (for a detailed analysis of this issue, see also Ref.~\cite{Li:2021cue}) since it contains a pole at $r=-a/2$, while Eq.~\eqref{ex:barX} is by construction free of this singularity and gives reasonable estimates for the compositeness. 
 Using $E_f=-7.2$ MeV and $g_1^2\ge 0.108$ for the $\chi_{c1}(3872)$, as discussed above , the $D^0\bar D^{*0}$ scattering length and effective range evaluated from Eq.~\eqref{Eq:ERE1ch} read $a = -28.6$ fm and $r \gtrsim -3$ fm. This yields $\bar X \gtrsim 0.9$ for the compositeness of the $\chi_{c1}(3872)$ from the data of Ref.~\cite{LHCb:2020xds}. Therefore, contrary to the claim of Ref.~\cite{Esposito:2021vhu}, this state is predominantly molecular. 

Similarly, using the ERE parameters from Eqs.~\eqref{Eq:aTcc} and \eqref{Eq:rTcc} one can estimate the compositeness of the $T_{cc}^+$ as follows. First, we remove the imaginary part from the inverse scattering length, 
since the unitarity contributions are additive for this quantity. 
Then, we remove the contribution from the higher-energy $D^+ D^{*0}$ channel to find 
the value $a= 10. 1$~fm for the scattering length in the $D^0 D^{*+}$ channel. 
Taking this value together with $r \gtrsim -11$ fm, where the range corrections were already included,\footnote{ While the scattering length in the LHCb analysis~\cite{LHCb:2021auc} was extracted from the scattering amplitude at the $D^0 D^{*+}$ threshold, the effective range was evaluated from the $K$-matrix and includes neither the contribution from the $D^+ D^{*0}$ channel nor the range corrections~\cite{Misha_privat}. } the compositeness of the $T_{cc}^+$ can be estimated as $\bar X \gtrsim 0.6$.
Thus, the ERE parameters extracted in Ref.~\cite{LHCb:2021auc} suggest that the $T_{cc}^+$ is not in conflict with the molecular scenario either, though more accurate information on the effective range is needed. 
This conclusion was further supported by the chiral EFT-based analysis of Ref.~\cite{Du:2021zzh}, where it was shown that the experimental data are consistent with a molecular interpretation of the $T_{cc}^+$ with the compositeness close to unity.

\section{Considering a finite width of a scattering state}
\label{sec:finitewidth}

It was pointed out in Refs.~\cite{Nauenberg:1962a,Braaten:2009jke} that the ERE needs to be modified
when one of the scattering particles is unstable. The reason for this is that the full amplitude
for the scattering of $h_1$ off $h_2$ with an allowed decay $h_1\to ab$ no longer has a two-body threshold
branch point at $M_2=m_{h_1}+m_{h_2}$, but a three-body branch point at $M_3=m_a+m_b+m_{h_2}<M_2$, and in addition
 two two-body branch points inside the nonphysical sheet at $m_{h_1}\pm i \Gamma_{h_1}/2 + m_{h_2}$~\cite{Doring:2009yv}.
 Accordingly, the range of convergence of the ERE is limited to $k\leqslant\sqrt{\mu\Gamma}\sim 7$ and 9~MeV for the $\chi_{c1}(3872)$ and $T_{cc}^+$, respectively,
 where for the estimate we used $\Gamma=\Gamma_{D^{* \, 0}}=55$ keV~\cite{Guo:2019qcn} and $\Gamma=\Gamma_{D^{* \pm}}=83$~keV~\cite{ParticleDataGroup:2020ssz}, respectively.
 To overcome this problem the authors of Ref.~\cite{Braaten:2009jke} suggest to replace the momentum in Eq.~(\ref{eq:ref})
 by
 \begin{equation}
k_{\rm eff} = \sqrt{2\mu (E+i\Gamma/2)} \ ,
\label{eq:keff}
 \end{equation}
which in practice implies that the expansion is performed no longer around the nominal (real) two-body
threshold but around the complex two-body branch point mentioned above. Note that the recipe of
Eq.~(\ref{eq:keff}) is approximate only, but in case of the $D^*$ the corrections which scale
as $\Gamma_{D^*}/(2\Delta)$, with $\Delta=M_2-M_3 \simeq 7$~MeV for the $\chi_{c1}(3872)$ case, are tiny~\cite{Hanhart:2010wh}.
Alternatively, one may fit the ERE to the scattering amplitude only for the
energies outside the range $|M_2\pm \Gamma|$,  that is,
excluding the energy range of $\pm \Gamma$ around the nominal
threshold where the threshold cusp, that would be prominent in the inverse amplitude for a stable $D^*$, in actuality
is modified by the finite $D^*$ width.
This is automatically done in experimental
analyses with the energy resolution worse than $\Gamma$.

\section{Summary and Conclusions}

The ERE provides a useful parameterisation of the scattering amplitude near the threshold. The lowest energy parameters, the scattering length and the effective range, 
are known to play a key role in determining the nature of a given state in the context of the Weinberg analysis of compositeness. However, care must be taken when
the interpretation of the effective range is considered in a coupled-channel problem. It is shown in this Letter that the appearance of a large and negative effective range, which 
in the one-channel case would indicate the dominance of a compact component in the wave function of a state, in a coupled-channel problem can be naturally generated by the 
coupled-channel hadron-hadron dynamics. We consider the two important examples of the $\chi_{c1}(3872)$ and $T_{cc}^+$ exotic states, treated as coupled-channel $D^0\bar D^{*0}/D^\pm\bar D^{*\mp}$ and 
$D^0D^{* +}/D^+D^{* 0}$ systems, respectively, and demonstrate that the effective ranges of these states acquire significant negative contributions driven by isospin violation in the masses 
of these two thresholds. In addition, we argue that the LHCb analysis~\cite{LHCb:2020xds} of the $\chi_{c1}(3872)$ line shape, with the current energy resolution, 
 can only give an upper limit on the absolute value of the effective range (in analogy with the LHCb analysis of the $T_{cc}^+$ in Ref.~\cite{LHCb:2021auc}), 
 while effective range values close to zero are also consistent with data. We estimate the finite-range corrections and the compositeness of the $\chi_{c1}(3872)$ and $T_{cc}^+$ states and conclude 
 that both states are not at all in contradiction with the molecular nature. Furthermore, the molecular component in the $\chi_{c1}(3872)$ is most likely the dominant one. 
Also, we discuss the modifications in the ERE in the presence of unstable constituents. 
 
\section*{Acknowledgments}
We would like to thank  Mikhail Mikhasenko and Sebasitian Neubert for useful discussions.
This work is supported in part by the National Natural Science Foundation of China (NSFC) and the Deutsche Forschungsgemeinschaft (DFG) 
through the funds provided to the Sino- German Collaborative Research Center ``Symmetries and the Emergence of Structure in QCD'' (NSFC Grant No. 12070131001, DFG Project-ID 196253076 -- TRR110), by the German Ministry of Education and Research under Grant No.~05P21PCFP4, by the NSFC under Grants No.~11835015, No.~12047503, No.~12125507 and No.~11961141012, by the Chinese Academy of Sciences under Grants No.~XDPB15, No.~XDB34030000, No.~QYZDB-SSW-SYS013 and No.~2020VMA0024, by the Spanish Ministry of Science and Innovation (MICINN) (Project PID2020-112777GB-I00), 
by the EU STRONG-2020 Project under the Program H2020-INFRAIA-2018-1, 
Grant Agreement No. 824093, and by Generalitat Valenciana under Contract PROMETEO/2020/023.
The work of A.N. is supported by the Ministry of Science and Education of the Russian Federation under Grant 14.W03.31.0026. The work of Q.W. is also supported by Guangdong Major Project of Basic and Applied Basic Research No.~2020B0301030008,
the NSFC under Grant No.~12035007, 
the Science and Technology Program of Guangzhou No.~2019050001, and
Guangdong Provincial Funding under Grant No.~2019QN01X172.


\end{document}